\documentclass[lettersize,journal]{IEEEtran}
\usepackage{amsmath,amsfonts}
\usepackage{algorithmic}
\usepackage{algorithm}
\usepackage{array}
\usepackage[caption=false,font=normalsize,labelfont=sf,textfont=sf]{subfig}
\usepackage{textcomp}
\usepackage{stfloats}
\usepackage{verbatim}
\usepackage{graphicx}
\usepackage{cite}
\PassOptionsToPackage{hyphens}{url}
\usepackage{hyperref}
\usepackage{url}
\usepackage[justification=centering]{caption} 
\begin{document}

\title{Mobile Edge Computing for the Metaverse}

\author{Chang Liu, Yitong Wang, Jun Zhao
        % <-this % stops a space
\thanks{The authors are all with Nanyang Technological University, Singapore. Email: JunZHAO@ntu.edu.sg}
}



\maketitle

\begin{abstract}
The Metaverse has emerged as the next generation of the Internet. It aims to provide an immersive, persistent virtual space where people can live, learn, work and interact with each other. 
However, the existing technology is inadequate to guarantee high visual quality and ultra-low latency service for the Metaverse players.
Mobile Edge Computing (MEC) is a paradigm where proximal edge servers are utilized to perform computation-intensive and latency-sensitive tasks like image processing and video analysis.
In MEC, the large amount of data is processed by edge servers closest to where it is captured, thus significantly reducing the latency and providing almost real-time performance.
In this paper, we integrate fundamental elements (5G and 6G wireless communications, Blockchain, digital twin and artificial intelligence) into the MEC framework to facilitate the Metaverse.
We also elaborate on the research problems and applications in the MEC-enabled Metaverse.
Finally, we provide a case study to establish a thorough knowledge of the user utility maximization problem in a real-world scenario and gain some insights about trends in potential research directions.

\end{abstract}


\section{Introduction}

The Metaverse is a computer-generated environment that fluidly combines physical reality with digital virtuality with a variety of consumer and business applications.
As a universal and immersive virtual world for users to create content, interact with each other and own digital assets, the Metaverse is believed to be the next iteration of the internet.
The potential of this internet evolution is attracting the attention of large groups of online game makers and social network companies~\cite{Bobrowsky2021BigTech,kraus2022facebook}, even making companies like Facebook pivot towards the Metaverse to stay competitive.
Unlike typical video games, which are entirely contained within a single ecosystem, the Metaverse is an open world for users to coexist.
To experience a different existence in virtuality that is a metaphor for the users' actual realities, each individual user in the Metaverse has her/his own avatar, which functions similarly to the user's physical person.
The technological enablers, particularly in the areas of mobile edge computing (MEC) and augmented and virtual reality (AR/VR) make the launch and quick uptake of the Metaverse platform possible~\cite{lee2021all}.

\textbf{Mobile Edge Computing.} Mobile Edge Computing (MEC), as an emerging computing paradigm, efficiently tackles the stringent constraints of communication and computation. Compared with conventional computing paradigms especially cloud computing, MEC proposed by the European Telecommunications Standard Institute (ETSI) introduces a novel strategy of task offloading. Instead of delivering the tasks to the central cloud servers, computation-intensive tasks of MEC are offloaded to the MEC servers that are placed on the edge of networks (e.g. base stations), which has a significant impact on reducing transmission delays and energy consumption of devices.

\textbf{Metaverse.} Facebook's rebranding of its company as ``Meta" and Omniverse developed by NVIDIA in the Metaverse system mark the next step in the Metaverse's evolution into the spotlight. This emerging ecosystem of services could significantly improve individuals' daily life in the virtual universe in terms of education, industries, and healthcare, where the inconvenience of living with an epidemic quarantine caused by COVID-19 can be tackled by the service applications in the Metaverse. As a universe of ternary-world interactions (i.e. physical world, human world, and digital world), Metaverse can provide a fully immersive world for people as digital avatars to interact between the physical and virtual worlds. More importantly, hyper spatiotemporal, self-sustaining, and interoperability are the inherent attributes of Metaverse as a result of the convergence of digital twin (DT), virtual reality (VR), augmented reality (AR), blockchain, and non-fungible token (NFT).

% \textbf{Play to Earn.} Play to Earn (PtL) blabla

\textbf{MEC-enabled Metaverse.} The convergence of mobile edge computing and the Metaverse meets both user-side and system-side requirements. For the user side, the edge servers close to the users are able to respond faster to the tasks sent by the devices. As Metaverse is based on a mass of virtual scenes and sensor data transmission based on mixed reality, mobile edge computing is able to meet the latency requirements of Metaverse service applications and solve the symptoms such as dizziness and vomiting of users due to high latency. Context-aware immersive content can also be delivered to improve the experience of users. Specifically, context-aware contents are distributed and cached in MEC servers based on the current users' contextual preferences, such as current fields of views (FOVs), view angles, and interacting actions. For the system side, MEC offloads tasks that would have been processed locally to the edge server, solving the problem of insufficient computing resources and battery life on local devices. A large amount of 2D/3D scene data from Metaverse applications can be cached in the edge server, solving the limitations of local device caching storage while reducing the total bandwidth required for immersive streaming and interaction.

\textbf{Comparison with other work.}
There are several studies on the Metaverse communication paradigms design and review.
Yu \emph{et al.}~\cite{yu20226g} provide a review on 6G mobile-edge empowered Metaverse, investigating the real-world deployment contexts of these technologies and proving that MEC with intelligent reflecting surfaces (IRS) can improve the energy efficiency and throughput in the Metaverse.
However, their study focuses on the 6G networks and enabling technologies. Our paper places emphasis on the MEC technologies that can empower the Metaverse.
Musamih \emph{et al.}~\cite{musamih2022metaverse} discuss the opportunities of the Metaverse in healthcare such as medical education and training, medical marketing and healthcare supply chain, as well as the enabling technologies.
In contrast, our paper provides insights into several Metaverse applications and enabling technologies.
The paper~\cite{aloqaily2022integrating} proposes a system that combines digital twins (DT) with other cutting-edge technologies, including the 6G communication network, blockchain, and AI. 
It focuses on sustaining continuous end-to-end Metaverse services, while our paper gives an overview of the enabling technologies, applications and research problems in MEC-enabled Metaverse.
% Liu \emph{et al.}~\cite{liu2022parallel} presents the definitions, applications, and current difficulties of parallel driving.

\begin{figure*}[tb]
\centering
\includegraphics[width=0.8\textwidth]{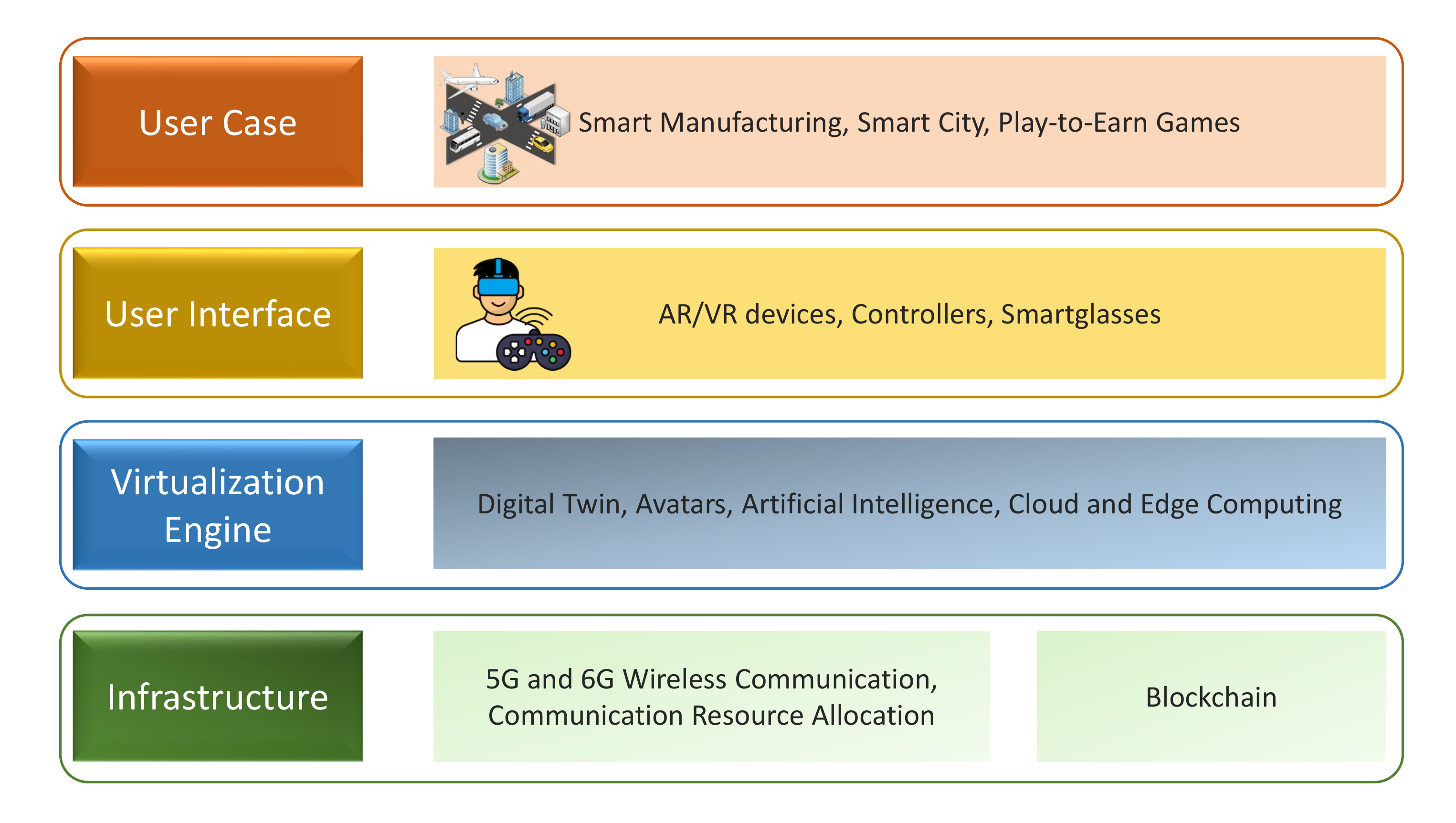}  
\caption{The framework of MEC-enabled Metaverse system.} \label{fig_arc}
\end{figure*}

\begin{figure*}[tb]
\centering
\includegraphics[width=0.9\textwidth]{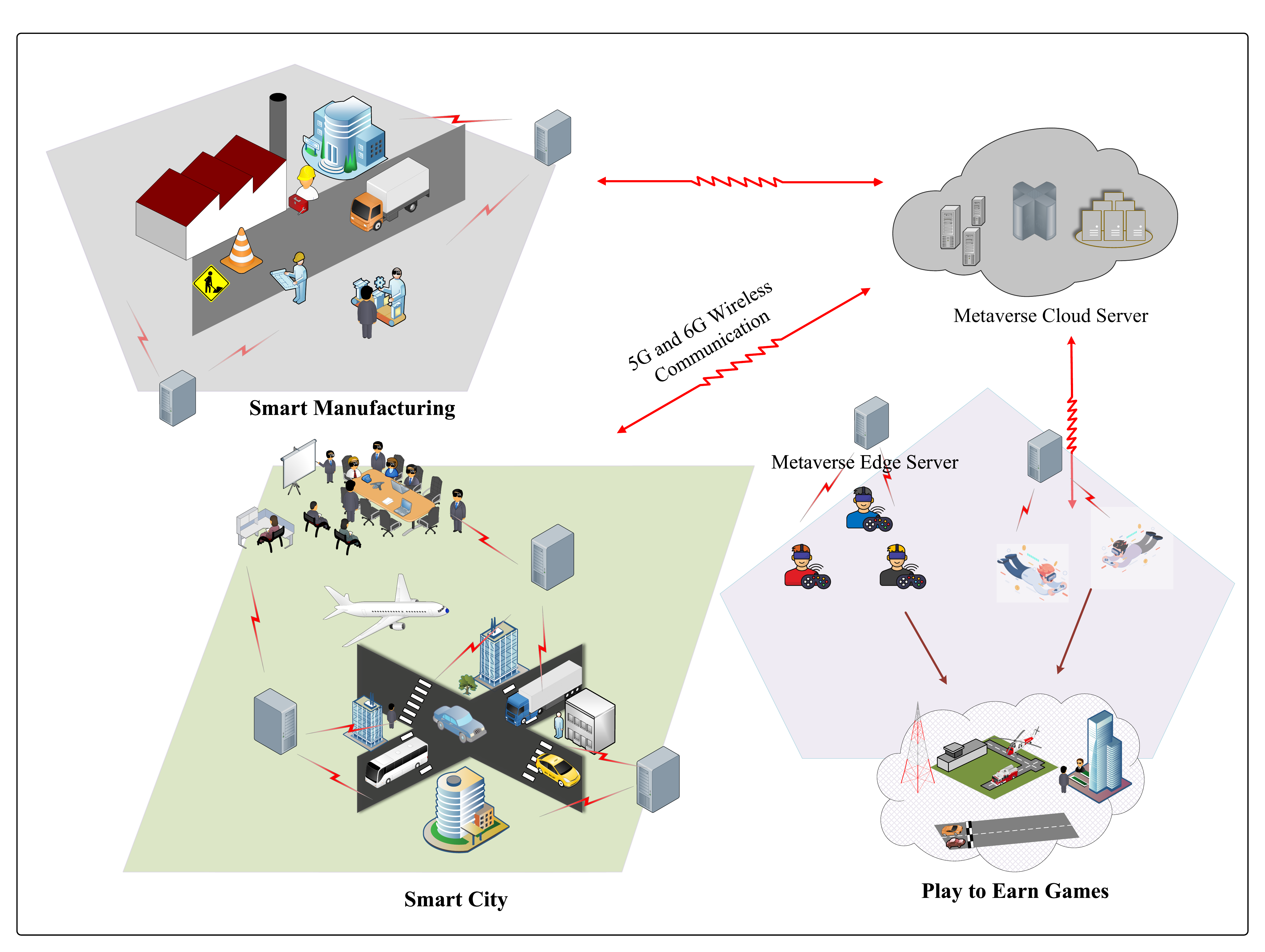}  
\caption{The applications of MEC-enabled Metaverse.} \label{fig_intro}
\end{figure*}
\section{Enabling technologies for MEC-enabled Metaverse}

In this section, the technologies that empower the Metaverse with MEC are discussed.
We give an overview of how 5G and 6G wireless communication, blockchain, digital twin and artificial intelligence can be developed to support MEC-enabled Metaverse.
Fig~\ref{fig_arc} illustrates the framework of the MEC-enabled Metaverse system. The wireless communication technologies and blockchain lay the infrastructure for the Metaverse system.
The virtualization engine includes digital twin and artificial intelligence.
Mobile devices such as AR/VR equipment serve as the user interface.

\subsection{5G and 6G Wireless Communication}
In contrast to the conventional 5th wireless communication network, the emerging 6G network is critical to meet the constraints in terms of delay, reliability, and application scenarios. Detailed comparisons are as follows: \par
5G: Based on the 5G attributes standardized by the Internet Telecommunication Union (ITU), three main 5G-based application scenarios are derived, including eMBB (enhanced Mobile Broadband), URLLC (Ultra Reliable Low Latency Communications), and mMTC (Massive Machine-Type Communications). 5G, as a new type of network, is the generation of the global wireless standard after 4G, designed to increase data transmission speeds, reduce latency and support more users, devices and services, while improving network efficiency. However, with the proliferation of communication devices and communication data volumes, as well as the evolution of communication scenarios, 5G will likely no longer meet some of the stringent requirements. 

6G: As distinct from 5G, the most inherent attribute of 6G is mobile broadband reliable low latency communication (MBRLLC)\cite{saad2019vision}. 6G, as a mobile communication system featuring ``connected intelligence", will provide more complete network coverage and a more secure communication network. In addition, 6G offers significant improvements in various metrics, such as a peak transmission rate of 1Tbit/s, which will enhance the performance of existing and future communication applications. Furthermore, holographic communication and high-fidelity human-computer interaction scenarios are further driving the rapid development of 6G. In short, the vision of 6G is to provide users with a more diverse and intelligent communication network based on realizing semantic communication\cite{strinati20216g}.

\subsection{Blockchain}
As a representative technology of decentralization, Blockchain can secure data of users in the Metaverse, improve transparency of transactions and prevent single point of failure (SPoF). The convergence of Blockchain and Metaverse can further improve the scalability and interoperability by using sharding technology and cross-chain technology to divide the whole Metaverse into multiple shards~\cite{xu2022full}. At the same time, Blockchain plays a key role in edge resource allocation by providing access control and identification of various services in the virtual world. In conclusion, Blockchain, as a key technology in the edge universe, can influence the improvement of various performances in the Metaverse, including computing offloading, resource allocation, and edge intelligence.

\subsection{Digital Twin}
Digital twin is a vital technology for projecting real-world scenes, objects and users into the virtual world. Specifically, digital twin, which makes full use of data such as physical models, sensor updates and operational histories, integrates multi-disciplinary, multi-physical quantity, multi-scale and multi-probability simulation processes and completes the mapping in virtual space, thus reflecting the full life-cycle process of the corresponding physical equipment.

\subsection{Artificial Intelligence}
Rendering the various scenarios of the large-scale Metaverse is inseparable from artificial intelligence (AI) technology. Specifically, the multilingual environment and personalised application services in the Metaverse are offered on the basis of massive data inference. AI technology also has a crucial role in enhancing the performance of the Metaverse system. For example, to meet variable and previously unknown time budgets, deep neural network (DNN) can be applied in the vision transformer backbones to improve the accuracy and speed of the system. Deep learning algorithm can also be adopted to identify the type (faithful or malicious) of edge servers to improve the security and convergence of the state-of-the-art systems\cite{9767623}.

\section{Research problems in MEC-enabled Metaverse}
Although the tremendous development of the above mentioned technologies, some problems still occur when large-scale MEC-enabled Metaverse system is deployed in the real world. In this section, we will identify and discuss the research problems in MEC-enabled Metaverse.

\subsection{Resource allocation}
Optimal allocation schemes of resources not only substantially enhance the computational performance of the system but also enable adaptation to new Metaverse service applications in the future. Resource allocation problems can be divided into two categories: communication resource allocation and computation resource allocation. For communication resource allocation, how to allocate FDMA bandwidth resources, TDMA time resources, and NOMA power resources emerges as the main concern, where FDMA, TDMA, and NOMA are short for frequency-division, time-division, and non-orthogonal multiple access, respectively. In particular, NOMA power allocation has gained more attention due to the fact that NOMA is the preferred alternative technology for 5G. For computation resource allocation, offloading decisions have a significant impact on it. Due to the different computing resources of VR devices and edge servers, computing tasks locally and offloading tasks to the edge servers are two preferred methods. But both approaches need to address the tradeoff between latency and resources. Thus, the novel device-edge association allocation of edge computation resources is progressively gaining focus in research.

\subsection{Mobility}
High fidelity and smooth scenes as users change their field of views (FOV) and physical position are essential factors in maintaining an immersive experience for users. However, seamless and stable services are subject to resource allocation and traditional optimization will have much computational complexity. To address mobility, deep reinforcement learning can be used and provides seamless services. For example, the prediction model uses past head movements to predict the users' position in the 360 degree view, which accurately provides the users with the next view of the screen. Furthermore, predicting the users' future location based on their physical motion can facilitate the advanced allocation of various resources.

\subsection{Security}
The security and privacy issues of the Metaverse are two issues that need to be addressed since the Metaverse is built on blockchain-empowered economic ecosystems~\cite{lim2022realizing}. For security issues, the following aspects are described. \textit{Data security}, as users trade goods, shares and other virtual funds in the virtual world, their virtual assets and transaction passwords must be secured from potential attackers. In addition, the user's activities in the Metaverse must be protected from hackers by the transmission of communication data and feedback from the server. Therefore, better authentication schemes and post-quantum cryptography are the dominant techniques to resist attackers. \textit{User security}, in the virtual world, collaboration among attackers will lead to digital avatars in the Metaverse receiving physical harm, for example, robbery, rape, etc. In the real world, attackers can commit data theft from the users' devices, leading to the prying of information about their real-world habitat and belongings.

\subsection{Privacy}
With regard to the privacy of users, service providers collect various human biological data from users to personalise their service provision, which can further lead to privacy breaches. For example, the users' eye information, including pupil spacing and pupil size, can be obtained by collecting eye tracking data. The users' sensitivity to colour and colour blindness can be determined by the users' reaction time to images of different colours. The users' nationality and native language can be identified by the user's reaction time to different languages. The performance and price range of the devices worn by the users can be judged by the RTT for different resolution animations. This data may be vital for companies to derive the attention span of users in order to market their products better, but if leaked, the users' personal privacy will also be stolen, or worse, attackers will steal or fake the users' identities to live in the Metaverse\cite{nair2022going}.\par
As a result, a number of techniques and risk assessment architectures to resist attackers should be pursued to defend users against the various threats they receive. The convergence of federated learning and homomorphic encryption can effectively improve the robustness of the system while protecting the privacy of the user.

\section{Applications of MEC-enabled Metaverse}
The Metaverse has a wide range of applications in the society and in the industry as illustrated in Fig~\ref{fig_intro}. In this section, we describe several Metaverse applications with great potential. 

\subsection{Smart Manufacturing}
The development of the Metaverse has not only taken people's daily lives from the real world to the virtual world but has also brought various industrial manufacturing activities into the virtual world. With the help of various VR/AR devices, people can even perform industrial activities that cannot be done in the real world, thus realising the ``industrial Metaverse". Specifically, professional surgeons can wear Microsoft Hololens 2 to perform preoperative simulations of high-risk surgical procedures and analyse any possible adverse effects on the patient during surgery. Automotive parts manufacturers can wear Microsoft Hololens 2 to perform the manufacturing and commissioning of high precision parts within the Metaverse, which not only saves the cost of failed part manufacturing but also maximises the performance of vehicles in the physical world.

\subsection{Smart City}
The form of the smart city of the future will gradually change into a virtual city. The Metaverse platform is an alternative to the smart cities of the future, enabling various social activities such as healthcare, economy, education and entertainment in a virtual world based on the convergence of various technologies, including Internet of Things (IoT), Blockchain, AI, VR, and AR\cite{allam2022Metaverse}. Yet, before the virtual city can be realised, the issue of how to plan the functioning of the city and the rights and interests of the people involved will have to be considered. For example, people can use AR technology to make furniture choices and match decorating styles and tones through AR devices. People will also be able to hold business and academic meetings in Metaverse while communicating face-to-face in a virtual world, solving the drawback of currently only being able to hold meetings on a 2D plane.

\subsection{Play-to-Earn Games}

With the aid of revolutionary technologies such as Blockchain, a great change is taking place in the game industry. 
The decentralized nature of the Metaverse allows players to gain real-world rewards from the time and effort they put into the game~\cite{lee2021all}.
This evolution of the game industry is known as \textit{``Play to Earn''} (P2E). In P2E games, players can log in to complete the game or battle with other plays to earn or create digital assets.
The most popular digital assets are cryptocurrencies and Non-Fungible Tokens (NFTs). 
In decentralized marketplaces, players are allowed to trade the NFTs with others in the game, and carry them to other digital experiences.
The brand-new, free-market and internet-based economy makes it increasingly prevalent that people spend more time on digital activities.
\vspace{-5pt}
\section{A Case Study of MEC-enabled Metaverse}

In this section, we will provide a case study of MEC-enabled Metaverse.

Mobile and wearable technology, such as AR glasses, headsets, and smartphones, is now the most appealing and popular Metaverse interfaces because they enable easy player movement.
However, the Metaverse applications' real-time interactive nature and their high demands on a seamless experience
are too heavy for mobile devices on account of the following two aspects:

\begin{itemize}
    \item \textbf{Limited Computation Resource:} To render a digital world which is physically accurate in the Metaverse, massive computation resources will be consumed.
    For instance, operating a typical AAA game today (such as Fortnite) takes multiple trillion floating point operations per second, and the need to generate completely immersive Metaverse experiences is anticipated to increase by two orders of magnitude.
    Even the most fundamental image and video processing tasks, such as real-time object detection, can become quite computationally complicated and time-consuming when used in increasingly sophisticated virtual environments~\cite{zhao2019object,cai2022compute}.
    However, the computational capacity of graphics and chipsets in the mobile devices such as AR glasses and headsets is not enough to handle sophisticated tasks. The most advanced all-in-one VR hardware, Oculus quest 2, is able to deliver VR service without a linked PC with its strong Qualcomm Snapdragon XR2 chipset~\cite{Qualcomm2020}. 
    The standalone VR experience comes at the expense of lower frame rates and hence less detailed VR sceneries because its computing power is still limited when compared to a PC or edge server.
    As a result, the user experience might suffer from insufficient processing resources if user activities incur delay~\cite{lee2020ubipoint}.
    
    \item \textbf{Limited Communication Resource:} The Metaverse is developed to support user interactions within the virtual world as well as with other users.
    To this end, it requires keeping track of the dynamic state information about physical devices and the integration of live streams and digital assets, as well as the dissemination of the processed streams.
    Specifically, in the Metaverse, the raw data collected by the sensors on the user side is transmitted to the server, and the processed information will be transmitted to the users again.
    The substantial amount of data that needs to be transmitted increases the network traffic between Metaverse users and the servers.
    For example, the video traffic in the Metaverse, already makes up 80\% of all Internet traffic today and is expected to expand at a 37\% compound annual growth rate over the next ten years, exceeding the present data use by about 24 times~\cite{Metaverseguide2022}.
\end{itemize}

The practice of offloading the computationally difficult work to a cloud server is a widely-used solution to the problem of mobile devices' limited computing resources.
A Metaverse Service Provider (MSP) edge server will often communicate the gaming environment in graphical form to the players in a typical Metaverse game scenario.
The players will take in the visual data and react. Players' states will alter in response to the player's appropriate actions. The MSP edge server will then receive these modifications and process them there.
However, due to the limited communication resource, transmission delay between the players and MSP edge servers will be increased when high-definition graphics data is transmitted. 
In the Metaverse applications over MEC, communication effectiveness is essential to provide players with an immersive gaming experience.
Virtual objects might move out of place if there is a significant amount of latency, which could affect players' reactions and the outcome of the profits they make in the game as a whole. This delay may also negatively affect players' health and dizziness, which will affect how they play the game.

Intuitively, the delay can be mitigated if one could decrease the size of transmission data.
Graphics will be of lower quality nevertheless if data is scarce. It might also hinder visuals, which would affect the gaming experience and cause difficulties in detecting the game's settings. 
In this case, players are more likely to play poorly, which results in fewer game tokens. As shown above, there is a clear trade-off between the amount of tokens a player might be able to earn and the volume of data being exchanged.
A further method to decrease service latency and boost token profits is to optimize the user-server association approach in addition to optimizing the data size communicated between player devices and edge servers. The varied geographic distribution of users often causes workload imbalances between edges, which leads to certain edge servers being overused and others being underused. Due to this imbalance, the overall performance of the entire Metaverse service system is reduced. It is crucial to develop a user-server association strategy in order to address this issue.
\begin{figure}[tb]
\includegraphics[width=0.5\textwidth]{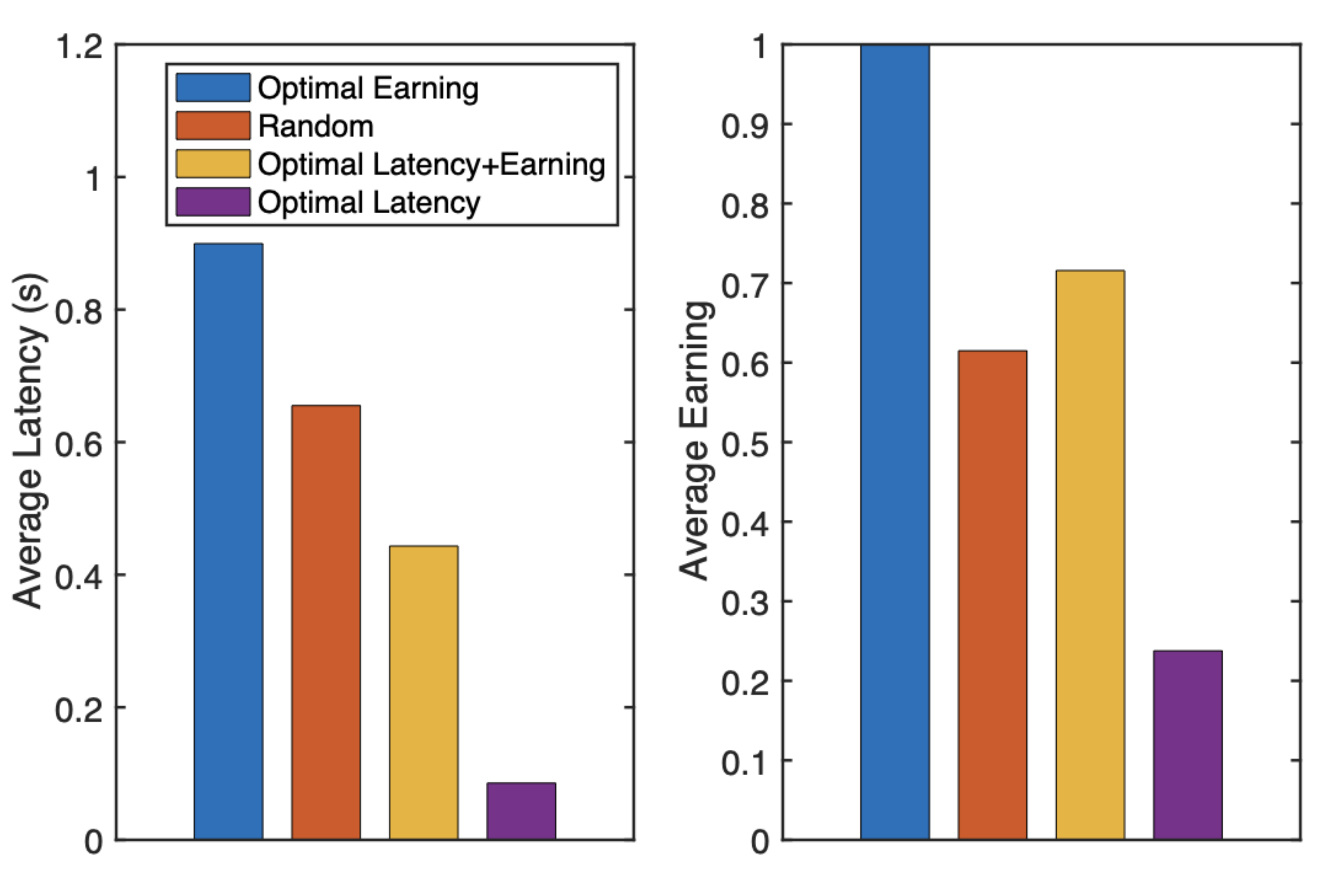}  
\caption{The latency and earning when weight parameter $\omega=2$.}
\vspace{-15pt}
\label{experiment_1}
\end{figure}

In this case study, we formulate an optimization problem to optimize the transmission data size by controlling the resolution, and the user-server association.
First, in order to measure the profits and describe how the service can affect the token earnings a player can make, we propose an increasing and concave resolution-impacted earning function. Through this function, a larger transmission data size will increase the visibility of the player, thus bringing higher profits to the player. If the transmission data size decreases, visual experience will be degraded, which might provide particular difficulties to the player, including making sense of the gaming scene. In this case, the tokens the player can earn are supposed to decrease. However, with the concave nature of the function, the marginal revenues will decrease when more data is transmitted.

To achieve a balance between the profits players can make and the latency players experience, we propose to maximize the utility of the players, which is defined as the weighted sum of profits and latency as follows:
\begin{align}
    \text{Utility} = \text{Profits} - \omega \cdot \text{Total Latency}.\nonumber
\end{align}
The weight parameter $\omega$ can be adjusted according to different preferences on latency. It allows us to capture the trade-off between the profits and the latency. The user-server association variables are integers while the transmission data size can be continuous.
Thus, the problem is a mixed-integer nonlinear programming which is considered to be NP-hard.
In addition, the nonlinearity in the objective function makes the problem hard to solve.
In order to tackle the problem, we propose to use alternating optimization to optimize the transmission data size and user-server association iteratively. 
For the transmission data size optimization, it is proved to be a convex optimization problem which can be solved by standard convex optimization software such as CVX in Matlab.
For the user-to-server optimization, it is a non-convex problem which can not be solved by convex optimization tools.
Thus, it is transformed into a quadratically constrained quadratic programming (QCQP) and we utilize the semidefinite relaxation (SDR) technique to solve the user-to-server problem and get a sub-optimal solution.
To verify the performance of the proposed algorithm, we simulate a system with 20 MSP edge servers and 100 Metaverse players.
A uniform distribution of Metaverse players' uplink transmission speeds ranges from 1 to 5 Mbps, while a uniform distribution of their downlink transmission rates ranges from 10 to 20 Mbps.
By measuring the latency with regard to different weight parameters $\omega=2$ and $\omega=4$, we evaluate the proposed methodology. The magnitude of latency's impact on the system is indicated by the value of $\omega$. The overall profits of users are then assessed for various minimum downlink data sizes.
We compare the proposed algorithm with three different algorithms: the ``Optimal Earning'' targets at making maximum profits while ignoring the players' delay experience. 
The ``Optimal Latency''  minimizes only the latency to acquire the ideal delay experience. The ``Random'' algorithm chooses random user-to-server association and transmission data size.
Our proposed algorithm is presented by ``Optimal Latency+Earning''. The average earning amount is scaled to fall between 0 and 1.

In Fig.~\ref{experiment_1}, the weight parameter $\omega$ is set to be 2. It can be seen that the ``Optimal Earning'' achieves the maximum profits at the expense of the highest latency. The ``Optimal Latency'' strategy maintains the lowest latency while acquiring little token earning.
Compared with ``Optimal Earning'', the proposed ``Optimal Latency+Earning'' reduces the latency by 50\% while obtaining 75\% of its token earning.
In comparison to ``Random'', the proposed strategy reduces the latency by 30\% and increases the earning by 16\%. 
\begin{figure}[tb]
\includegraphics[width=0.5\textwidth]{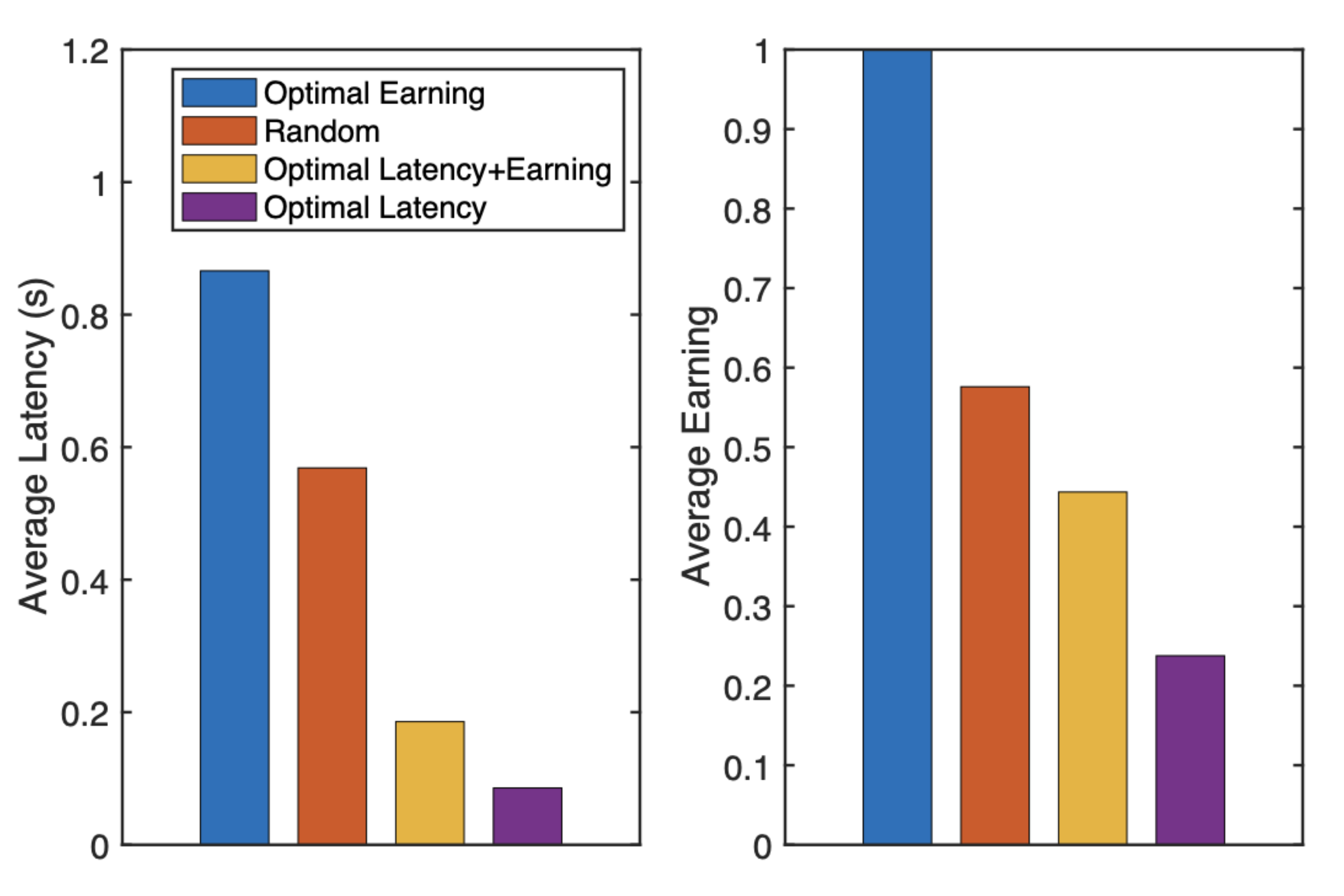}  
\caption{The latency and earning when weight parameter $\omega=4$.}
\vspace{-15pt}\label{experiment_2}
\end{figure}
In Fig.~\ref{experiment_2}, the weight parameter $\omega$ is specified to be 4. It indicates that more attention is placed on latency. 
It can be observed that our technique generates fewer profits when $\omega=2$. It means the system exchanges latency for profits. The ``Optimal Latency'' approach maintains low latency continuously since it only seeks to reduce total latency. 
As a result, it consistently picks the smallest downlink data size and earns little tokens. The proposed ``Optimal Latency+Earning'' achieves only 36\% latency of ``Random'' strategy but obtains 81\% of its token earning.
\vspace{-15pt}
\section{Discussions}
The deployment and application of the Metaverse are still at an early stage. A large number of practical problems are there when large-scale Metaverse applications are developed. 
Many challenges come from the high requirement on user-oriented experience with limited network resources and the heterogeneity of different traffic patterns.
\vspace{-5pt}
\subsection{Hierarchical Edge-Cloud-Enabled Metaverse}
The network framework that is used widely today assumes a central cloud server as a Metaverse service provider where the delivery of service is easy. However, incorporating the concept of mobile edge computing, a three-layer hierarchical framework enables the edge server and cloud server to collaborate to provide Metaverse service.
Low bandwidth demands, data storage locally and low latency Metaverse applications are made possible by the edge servers. 
Although an edge-based system enjoys a low delay, the number of devices it can access is limited. When it involves a machine learning process, the limited number of participating devices will degrade the system's overall performance.
While the cloud server can access more devices to train a more accurate model, an edge-cloud-enabled hierarchical Metaverse system can achieve a balance of the trade-off between latency and learning performance.
Further studies are necessary on the design of an edge-cloud-enabled hierarchical Metaverse system, including how to design the offloading strategy and how to associate the players with edge servers and the cloud.
\vspace{-5pt}
\subsection{Metaverse with Augmented Reality (AR), Virtual Reality (VR) or Mixed Reality (MR)}

%different traffic patterns

Augmented reality (AR), virtual reality (VR) and mixed reality (MR) are all combined into the much larger idea of the Metaverse.
The phrase that most people are acquainted with is probably virtual reality. A 3D virtual environment created by a computer that either simulates the actual world or an invented one is referred to as VR. A physical environment is seen in real-time with its aspects enhanced by digital sensory input like music, images, films, or GPS information. In other words, AR overlays the physical world by enabling users to view the real world through the displays of their phones or tablets. MR, also known as hybrid reality, is the blending of the virtual and real worlds to create a new setting where real-time interactions between virtual and real-world items are possible.
Many unanswered concerns remain about the future operation of the Metaverse. Undoubtedly, AR, VR and XR are crucial to the development of the Metaverse.
It is necessary to facilitate the growth, in order to ensure long-standing connectivity and smooth virtual experiences for users.
\vspace{-2pt}
% As the business grows and customers become more affluent, the industry will assuredly get more accurate.

\section{Conclusion}
This paper provides an investigation on the MEC-enabled Metaverse. MEC utilizes the Metaverse edge server which is close to a particular data source to provide real-time data processing service. It could significantly reduce the latency, optimize the wireless communication bandwidth utilization and ensure an immersive user experience for Metaverse players.
This article gives an overview of the enabling technologies, research problems and MEC-enabled Metaverse applications.
A case study is provided where the transmission data size and user-to-server association are optimized to maximize users' utility.
Finally, the paper discusses some future research challenges and directions that may arise when a large-scale MEC-enabled Metaverse system is deployed in the presence of limited network resources and heterogeneous traffic patterns.
\vspace{-5pt}
% \bibliographystyle{IEEEtran}
% \bibliography{related}



\vspace{-15pt}
\section*{Biographies}
Chang Liu is currently pursuing a Ph.D. degree with School of Computer Science and Engineering, Nanyang Technological University. His research interests include federated learning and Metaverse.

Yitong Wang is currently pursuing a Ph.D. degree with School of Computer Science and Engineering, Nanyang Technological University. His research interests include mobile edge computing.

Jun Zhao (S'10-M'15) is currently an Assistant Professor in the School of Computer Science and Engineering at Nanyang Technological University (NTU) in Singapore. His research interests include blockchain, edge/fog computing, and Metaverse.

\vfill

\end{document}